\documentclass[12pt,a4paper]{article}
\usepackage[utf8]{inputenc}
\usepackage[english]{babel}
\usepackage{amsmath}
\usepackage{amsfonts}
\usepackage{authblk}
\usepackage{csquotes}
\usepackage{xcolor}
\usepackage{float}
\usepackage[colorlinks=true, linkcolor=blue, urlcolor=blue, citecolor=blue]{hyperref}
\usepackage{amssymb}
\usepackage{graphicx}
\usepackage[left=2.5cm,right=2.5cm,top=4cm,bottom=2.7cm]{geometry}
\usepackage{fancyhdr}
\newcommand{\D}{\mathop{}\!\mathrm{d}}
\newcommand{\half}{\tfrac{1}{2}}
\newcommand{\sqg}{\sqrt{-g}}
\newcommand{\dmu}{\partial_{\mu}}
\newcommand{\dnu}{\partial_{\nu}}

\newcommand{\nmu}{\nabla_{\mu}}

\newcommand{\nro}{\nabla_{\rho}}
\newcommand{\nsi}{\nabla_{\sigma}}
\newcommand{\Boxop}{\Box}

\title{An Analytic Formalism of Inflation for Derivative Coupled Scalar Field and Validating
its predictions for Some Inflationary Potentials}
\author{Aayush Randeep$^{1}$\footnote{aayush21@iiserb.ac.in}   and Rajib Saha$^{1}\footnote{rajib@iiserb.ac.in}$}
\affil{\small
$^1$ Department of Physics, Indian Institute of Science Education and Research, Bhopal, India
}

\date{}
\begin{document}
\maketitle
\begin{abstract}
\footnotesize
One of the fundamental objectives of contemporary cosmology is to understand the
physics of the inflationary universe, owing to its observably verifiable predictions about the very early universe with an energy scale of $\sim 10^{16}$ GeV. Recent observations from the
ACT and  the Planck mission, constrain the values of  the scalar spectral index, $n_s$, and the tensor-to-scalar ratio, with state-of-the-art accuracy and upper limits, respectively. In the current work, a type of non minimally coupled inflationary model in which the gravity and the background scalar field interact through a covariant product of the Ricci tensor and derivatives of the scalar field. With this interaction at the backdrop, we estimate $n_s$ and $r$
for a wide range of inflaton self-interaction potentials, including power law, exponential $\alpha$ attractor, Arctan, Hilltop, and polynomial model. We show that the higher derivative
terms involving the scalar field resulting from the derivative coupling term can be handled
without facing any singularity within the  slow-roll regime.  We show that it is possible to produce $n_s$ and
$r$ values consistent with ACT and Planck observations for each of the chosen sets of potentials for the
derivative coupled action.

\end{abstract}
{\footnotesize\tableofcontents}
\section{Introduction}
The epoch of Inflation \cite{Guth1981InflationaryUniverse}, marked by an accelerated expansion of spatial scales before the creation of particles predicted by standard particle physics, is a well-established phenomenon in Cosmology.  Although the phenomenon was first introduced to address the flatness, horizon, and monopole problems of the standard cosmology \cite{LiddleLyth2000}, which contain matter and radiation, it also provides a well-defined mechanism for the generation of quantum fluctuations, which form the basis for the formation of large-scale structures in the Universe at a much later epoch. A wide variety of models of inflation have been proposed since the inception of the basic idea. We refer to Refs  \cite{Martin:2013wta} \cite{Gomes:2018xgx} \cite{Vazquez:2018inflation} for a description of this development.

An interesting route to inflation is through the high-energy framings inspired by string theory, such as brane inflation and axions. They provide an explanation for the inflaton potential flatness problem through symmetries and ultraviolet completions, effectively addressing its high-energy nature. One of them is non-minimal derivative couplings (NMDC), which involves couplings between the kinetic term of the inflation field and curvature tensors.  Such theories were originally introduced as a natural generalisation of scalar tensor theories and, due to their seeming complexity and presumed instabilities for models involving high-order derivatives, largely abandoned in the late nineties \cite{Amendola1993}. However, a shift in the scientific community's interest occurred with the realisation that derivative couplings with the Einstein tensor preserve second-order field equations and support an improved gravitational friction mechanism \cite{GermaniKehagias2010}. This extra friction enables steep scalar potentials that fail to satisfy the slow-roll conditions for successful inflation, thereby greatly multiplying the parameter and model landscape. In view of ever-better CMB constraints on (high) scalar spectral index, with measured value implying near scale invariance (e.g., Planck satellite missions \cite{Planck2018} and ACT  \cite{calabrese2025atacamacosmologytelescopedr6}), NMDC models of inflation gained renewed attention due to their natural prediction for low tensor modes and spectral slopes fitting prevailing observations. This revived interest made NMDC models of inflation highly plausible as an alternative framework for slow-roll
models in inflation theory, inspired this time not by theories for possible modifications of gravity on cosmological scales or distances, but rather by their detailed matter-couplings
in spacetime geometries.

In general, a non-minimal derivative term  \cite{HrycynaSzydlowski2015} \cite{Yi_2016} \cite{CapozzielloLambiaseSchmidt1999} \cite{Gao2025NonMinimalDerivativeCoupling} can be regarded as a general case of scalar-tensor theory \cite{QuirosKumar2025} as it can be shown that a conformal transform maps it to a
scalar-tensor theory \cite{Gao2025NonMinimalDerivativeCoupling}.  However, if we have a term of the type in which a curvature
term is coupled with the kinetic part of the scalar field, then the former transformation is not possible. It also provides an additional parameter that we can tune. It provides a rather simple way to add a nontrivial part to the analysis. This approach has the benefit of defining all the parameters fundamentally. The additional constant, which we refer to as the coupling constant, also serves a phenomenological purpose. However, it can be complex to treat  \cite{CapozzielloLambiaseSchmidt1999}. We will demonstrate that in our case, we have a nicely behaved field equation. We will treat it with popular potentials and show that it agrees well with
the observation \cite{villagra2025atacamacosmologytelescopehighredshift}.  We have used the DR 6 \cite{louis2025atacamacosmologytelescopedr6} release of the ACT  \cite{villagra2025atacamacosmologytelescopehighredshift} observations to validate our predictions. It favours a higher spectral tilt. The latest data release from the Atacama Cosmology Telescope (ACT) provides strong support for the inflationary paradigm  \cite{ACT2023} \cite{ACTPol2023}. In addition to reinforcing the standard cosmological picture, the ACT measurements, when combined with large-scale structure information from the DESI DR1 data release  \cite{DESI2025_DR1} \cite{DESI2025_JointCosmo}, indicate a possible shift in the preferred value of the scalar spectral index $n_s$ compared to earlier cosmic microwave background–only analyses. The benchmark constraint on the spectral index was established by the Planck 2018 analysis, which reported
$n_s = 0.9651 \pm 0.0044$ \cite{Kallosh:2025chaotic}, and has since been widely used in tests of inflationary models  \cite{Planck2018}. Over the years, however, several analyses have pointed toward a mild upward trend in the inferred value of $n_s$. For example, Ref.~ \cite{HigherNsTrend} obtained $n_s = 0.9683 \pm 0.0040$, suggesting a modest deviation from the original Planck-only result.
This trend becomes more pronounced in joint analyses combining Planck, ACT, and DESI DR1 data. The combined P-ACT-LB analysis yields $n_s = 0.9743 \pm 0.0034$,which differs from the Planck 2018 value at approximately the $2\sigma$ level  \cite{ACT2023}. If this shift is confirmed, it may have important implications for inflationary model building, potentially disfavouring models that predict lower values of $n_s$ and motivating a re-examination of assumptions entering combined cosmological datasets.
 The motivation to add this NMDC term is rather natural. Keeping in mind the homogeneous and isotropic
scalar field, any additional term in the action will ultimately contribute to the derivatives of the scalar field \cite{DeFelice2010fR}. This provides a means to adjust the damping, which should result in a higher spectral tilt value as desired  \cite{GermaniKehagias2010}.  Even in that case, motivated by other NMDC models, we have taken the high-friction limit, which further enhances the damping  \cite{GermaniWatanabe2011}. The high-friction limit also helps in setting the strength of the coupling term, allowing the
two to fit together nicely  \cite{GermaniWatanabe2011}.  In the slow-roll assumption, it turns out that the coupling
corrections are not that different \cite{Gao2025NonMinimalDerivativeCoupling}. The reason behind this can also be understood in
terms of the slow-roll parameters. For example, the most popular NMDC model, which
utilises the Einstein tensor as the curvature term, aligns almost identically with ours due
to the first slow-roll parameter \cite{GermaniKehagias2010}.
In our analysis, we demonstrate that the addition of the Ricci Tensor as the NMDC
term indeed helps us achieve the desired higher value of the spectral tilt. We start
by minimising the action and obtaining the equations of motion. We then impose
the slow-roll conditions, obtain the slow-roll parameters, and subsequently do the power
spectrum analysis to compare with the observations \cite{louis2025atacamacosmologytelescopedr6}.

To validate the theoretical predictions, we then take several important potentials. The full
analytical calculation can be performed for the power-law potential. The rest, namely
Exponential, Hilltop, arctan, and polynomial attractors are treated numerically. The integration step to calculate the number of e-folds is the one step where the analytic calculation is rather difficult  \cite{LiddleLyth2000}.  The power-law potential and the arctan potential lie completely beyond the 1$\sigma$ region. All other show good compatibility with the observations as they all
lie within the 1 $\sigma$ region \cite{louis2025atacamacosmologytelescopedr6}. In future, we are also planning to add another modified
term to the NMDC term to furnish one more parameter that we can tune. It will be also
be interesting to see, how the theoretical observation change with other curvature terms,
even developing a general framework to accommodate a general tensor \cite{Horndeski1974}.

In future, we are also planning to add another modified
term to the NMDC term to furnish one more parameter that we can tune. It will be also
be interesting to see, how the theoretical observation change with other curvature terms,
even developing a general framework to accommodate a general tensor \cite{Martin2014Encyclopaedia} \cite{Silverstein2008EFT}  \cite{DeFeliceTsujikawa2010}. A natural  approach  is experimenting with different potentials \cite{Sousa2023OptimalPotentials}. The potential energy density of the scalar field plays an important role in determining the dynamics and the extent of inflation. The search for a suitable potential  \cite{Sousa:2023etp} has been an interesting topic of research. Other than that, there are modified gravity theories \cite{Starobinsky1980} \cite{DeFeliceTsujikawa2010} \cite{BransDicke1961} \cite{Horndeski1974} \cite{Nojiri2005GB}  that add
a small perturbation to the Einstein-Hilbert action  \cite{Einstein1916GR}. Aside from the traditional single
field kinetic inflation in the context of Einstein gravity, many exotic inflationary scenarios
have been constructed by extending the inflaton sector rather than gravity. These models
comprise non-canonical kinetic inflaton theories, such as k-inflation  \cite{ArmendarizPicon1999kInflation} and curvaton
scenarios  \cite{Lyth2002Curvaton}, which introduce new scalar fields in addition to the inflaton.

The discovery of CMB over the past decades has made it possible to test the inflationary theory with precision like never before. Beginning in the late ’90s, COBE \cite{Smoot1992COBE} \cite{Bennett1996COBE} \cite{Mather1994FIRAS} gave the first evidence that the CMB was a perfect blackbody and had a surprisingly uniform temperature. Later in 2000, WMAP \cite{Spergel2003WMAP} \cite{Komatsu2009WMAP} \cite{Komatsu2011WMAP} \cite{Hinshaw2013WMAP} provided a clearer picture, enabling us to obtain the angular power spectrum in the familiar form. The increased resolution helped us to estimate several important features of our universe, like the age, composition, solidifying the $\Lambda$CDM  \cite{Einstein1917Lambda} \cite{Planck2018LCDM} \cite{Peebles2003LCDM} model and the existence of dark matter  \cite{Rubin1980DM} \cite{Zwicky1933DM} and dark energy \cite{Riess1998DE} \cite{Perlmutter1999DE}. The natural model used was the scalar field  \cite{LiddleLyth2000}  to get the accelerated expansion. The equations of motion obtained are dealt with in the slow-roll regime \cite {Linde1982Inflation} \cite{LiddleLyth2000}, motivated by exponential expansion. The potential is assumed to be nearly flat, allowing the scalar field to roll slowly. The kinetic part, i.e., can then be negligible as compared to the potential term, simplifying the analysis significantly. These assumptions are captured by the slow-roll parameters, which will be seen later. They are framed such that, as long as they are smaller than unity, the inflation is in progress. However, it has to end, so the moment any of them becomes unity, inflation is said to have ended.
\section{Formalism}
\subsection{Action and Equation of motion}

We propose a model in which in addition to the canonical action of inflation the kinetic part is coupled to the Ricci Tensor via the coupling constant $1/\lambda^2$ (which has the  dimensions of mass squared inverse), forming the NMDC term. The action takes the following form,

\begin{equation}
S=\dfrac{1}{2} \int d^4x \sqrt{-g} \left[R-g^{\mu \nu}\partial_{\mu}\phi \partial_{\nu}\phi -\dfrac{1}{\lambda^2}R^{\mu \nu}\partial_{\mu}\phi \partial_{\nu}\phi-2V(\phi)\right].
\end{equation}
The action can be written in more suggestive form as sum of three components, 

\begin{eqnarray}
S_1 &=& \dfrac{1}{2}\int d^4x \sqrt{-g} R , \label{EH Action} \\
S_2&=&\dfrac{1}{2} \int d^4x \sqrt{-g} \;g^{\mu\nu} \partial_\mu \phi \partial_\nu \phi \; -2V(\phi) , \label{M Action} \\
S_3 &=& -\dfrac{1}{2\lambda^2} \int d^4x \sqrt{-g} R^{\mu\nu} \partial_\mu \phi \partial_\nu \phi,  \label{NMDC Action}
\end{eqnarray}    
where $S_1$ is the Einstein-Hilbert action \cite{Einstein1916GR} forming the gravity part of the canonical action and $S_2$ becomes the matter part. Varying the action  Eq. \eqref{EH Action} with respect to the metric gives the familiar Einstein tensor, Eq. \eqref{M Action} forms the Stress-Energy tensor and $S_3$ is the new NMDC term. The NMDC part in the action however, gives rise to two components. It will be shown that the first part is what goes into the dynamics, the latter part can be dropped within the consideration of slow-roll. The variation of each can be written as follows,

\begin{equation} \label{eh minimised}
    \delta S_1 = \dfrac{1}{2} \int d^4x \sqrt{-g} \; \delta g^{\mu \nu} \left[ R_{\mu \nu}-\dfrac{1}{2}Rg_{\mu \nu}\right] ,
\end{equation}

\begin{equation} \label{matter minimised}
    \delta S_2 = \int d^4x \sqrt{-g} \; \delta g^{\mu \nu}  \left[\partial_\mu \phi \partial_\nu \phi -\dfrac{1}{2}\left(g^{\alpha\beta} \partial_\alpha \phi \partial_\beta\phi+ V(\phi)\right) \right],
\end{equation}

\begin{equation} \label{S3}
    \delta S_3 =  -\dfrac{1}{2\lambda^2}\int d^4x \sqrt{-g} \; \delta g^{\mu \nu} \left[\dfrac{1}{2} R^{\alpha\beta} \partial_\alpha \phi \partial_\beta\phi g_{\mu\nu}\right] -
    \dfrac{1}{2\lambda^2}\int d^4x\;\sqrt{-g}\; \left[\delta R^{\mu\nu}\partial_\mu \phi \partial_\nu \phi\right].
\end{equation}
The coupling contributes two terms. The first term is obtained by varying the determinant of the metric, the second part involves the variation of Ricci tensor. We write $\delta S_3$ as a sum of two variations $\delta I_1$ and $\delta I_2$,

\begin{equation}
\delta S_3 = \dfrac{1}{2 \lambda^2} \left(\delta I_1 + \delta I_2 \right),
\end{equation} 

\begin{equation}
\delta I_1 = -\dfrac{1}{2} \int d^4x\; \sqrt{-g} \;\delta g^{\mu \nu} \left[R^{\alpha\beta} \partial_\alpha \phi \partial_\beta\phi g_{\mu\nu}\right], 
\end{equation}

\begin{equation} \label{higer_order}
    \delta I_2=- \int d^4x\;\sqrt{-g}\; \left[\delta R^{\mu\nu}\partial_\mu \phi \partial_\nu \phi\right].
\end{equation}
In order to evaluate $\delta I_2$ we invoke the Palatini identity which serves as a bridge to map the variation of the ricci tensor to the variation of the metric. We also require the variation of the Christoffel symbols both of which are stated below,

\begin{equation}
    \delta R_{\mu\nu}
= \nabla_\lambda (\delta \Gamma^\lambda_{\ \mu\nu})
 - \nabla_\nu (\delta \Gamma^\lambda_{\ \mu\lambda})  , \label{palatini}
\end{equation}

\begin{equation}
  \delta\Gamma^{\lambda}{}_{\mu\nu}
  = \half g^{\lambda\sigma}\!\left(
      \dmu\,\delta g_{\nu\sigma}
      + \dnu\,\delta g_{\mu\sigma}
      - \partial_{\sigma}\delta g_{\mu\nu}
    \right).
  \label{eq:palatini}
\end{equation} \\[5pt]
Substituting the Eq. \eqref{palatini} into $\delta I_2$ in Eq. \eqref{higer_order} and performing integration by parts on each of the terms we obtain the following form,

\begin{equation}
\delta I_2 = -\int \sqrt{-g} \, (\delta \Gamma^\lambda_{\ \mu\nu}) 
  \nabla_\lambda (\partial^\mu \phi \, \partial^\nu \phi)\, d^4x
+ \int \sqrt{-g}\, \delta\Gamma^\lambda_{\ \mu\lambda}\,
  \nabla_\nu(\partial^\mu \phi \partial^\nu \phi)\, d^4x.
\end{equation}
Substituting Eq. \eqref{eq:palatini} into the first term of $\delta I_2$ and noting that  $\nabla_{\lambda}(\partial^{\mu}\phi\,\partial^{\nu}\phi)$ is symmetric in $\mu,\nu$, after integration by parts on each of the terms. The second part of $\delta I_2$ can be handled on the similar lines. Adding both the parts of $\delta I_2$ we obtain ,
\begin{equation}
  \delta I_2
  = \int \sqg\; \D^{4}x\; \delta g^{\mu\nu}
    \left[
      -\nro\nmu\!\left(\partial^{\rho}\phi\,\dnu\phi\right)
      +\half\Boxop\!\left(\dmu\phi\,\dnu\phi\right)
      +\half g_{\mu\nu}\,\nro\nsi\!\left(\partial^{\rho}\phi\,\partial^{\sigma}\phi\right)
    \right].
  \label{eq:deltaI}
\end{equation}
Summing all three terms, we see that at least in the first friedmann equation the third order term is cancelled giving us no contribution. In the slow-roll case, we could have skipped the calculations starting from the equation Eq. \eqref{eq:deltaI}. Each of the terms becomes proportional to the factor of 
$\ddot\phi^{2}+\dot\phi\dddot\phi$. The contribution to the field equation from Eq. \eqref{EH Action} and Eq. \eqref{M Action} are both proportional to $\dot{\phi}^2$, hence comparing the terms we have the ratios 
\begin{equation}
\frac{\ddot{\phi}^{\,2}}{\dot{\phi}^{\,2}}
 + 
 \frac{\dddot{\phi}}{\dot{\phi}} \;,
\end{equation} which is obviously negligible in the case of slow-roll. The point here is for any other non-trivial tensor the following cancellation might not work out to be exaclty zero, but the above argument will still hold. However,the $ij$ contribution from each term does not posses this lucky cancellation, 
The first term evaluates to zero: $-\nabla_{\rho}\nabla_{i}(\partial^{\rho}\phi\partial_{j}\phi) = 0$. This is because it only has a 00 component on the FRW background. The second term also evaluates to zero: $+\frac{1}{2}\Box(\partial_{i}\phi\partial_{j}\phi) = 0$. This evaluates to 0 because spatial gradients vanish on a spatially flat FRW background. Since the product of the gradients is 0, its D'Alembertian is identically 0. The third term evaluates to $+\frac{1}{2}g_{ij}\nabla_{\rho}\nabla_{\sigma}(\partial^{\rho}\phi\partial^{\sigma}\phi) = +a^{2}\delta_{ij}(\dot{\phi}^{2}+\dot{\phi}\ddot{\phi})$. The final form of the Friedmann equations take the final form,

\begin{equation} \label{first eom}
H^2=\dfrac{1}{3}\left[\dfrac{\dot{\phi^2}}{2}\left(1+\dfrac{3K}{\lambda^2}\right)+V(\phi)\right],
\end{equation}

\begin{equation}
    2\dot H + 3H^{2}
    = -\frac{\dot\phi^{2}}{2} + V(\phi)
      - \frac{1}{2\lambda^{2}}\!\left(\ddot\phi^{2}+\dot\phi\dddot\phi\right).
  \label{eq:ij_Friedmann}
\end{equation}
The equation of motion can be obtained by varying the action with respect to the field using the Euler-Lagrange Equation \cite{Carroll2004} \cite{Wald1984} \cite{Weinberg1972} in curved spacetime.
\begin{equation} \label{second eom}
\dfrac{d}{dt}\left[a^3 \dot{\phi}\left(1+\dfrac{3K}{\lambda^2}\right)\right]=-a^3 V',
\end{equation} 
where $K=\dfrac{\ddot{a}}{a}$ 
It is evident that without the coupling terms, we reduce back to the canonical theory. We will do our analysis in the slow-roll regime  \cite{Liddle_1994}. The idea is that we add non-minimal derivative terms, which in the isotropic and homogeneous  \cite{Rudnicki1993} case ultimately contribute to the damping term. In the high-friction limit, we further enhance the damping, which physically means that the scalar field rolls slowly. This helps in attaining higher values of the spectral tilt.

\subsection{Slow Roll Conditions}
In this section, we will put slow-roll constraints on the equation of motion. 
Taking the derivative of equation Eq. \eqref{second eom}  and expanding, we have,

\begin{equation}
3H\dot{\phi}\left(1+\dfrac{3K}{\lambda^2}\right)+\ddot{\phi}\left(1+\dfrac{3K}{\lambda^2}\right)+\dot{\phi}\left(\dfrac{3\dot{K}}{\lambda^2}\right)=-V'.
\end{equation}
In slow-roll, the second time derivative of $\phi$ should be much smaller than the first time derivative, giving us the first slow-roll condition, which is in fact the same as it would have been without the coupling,
\begin{equation} \label{first slow roll}
\ddot{\phi}<<3H\dot{\phi} \;.
\end{equation} 
We note that there are two terms containing $\dot{\phi}$; however, $K$ is nothing but $\dot{H}+H^2$. Now, invoking the definition of the first slow roll condition in the $H$ basis, and comparing the second term, it can expressed in terms of the first slow-roll parameter as follows,
\begin{equation}
\left|\dfrac{\dot{K}}{H(\lambda^2+3K)}\right|= \left|\dfrac{2F \epsilon_H^2}{1+3F(1-\epsilon_H)}\right|,
\end{equation} 
where $F=H^2/\lambda^2$ and $\epsilon_H=-\dot{H}/H^2$. The RHS of the equality is second order in $\epsilon_H$, hence very small than unity, giving us our second slow roll condition. The third slow-roll condition follows trivially from Eq. \eqref{first eom}
\begin{equation}
\left|\dfrac{\dot{K}}{H(\lambda^2+3K)}\right| <<1  ,
\end{equation} 
\begin{equation}
	\dfrac{\dot{\phi^2}}{2}\left(1+\dfrac{3K}{\lambda^2}\right)<<V(\phi) .
\end{equation}
The construction is made with the understanding that, without the coupling, the theory should be reduced to the standard case. Under these conditions, we have equations Eq. \eqref{first eom} and Eq. \eqref{second eom} ,
\begin{equation} \label{first simplified eom}
H^2 \approx \dfrac{V(\phi)}{3} ,
\end{equation}  
\begin{equation} \label{second simplified eom}
3H\dot{\phi}\left(1+\dfrac{3K}{\lambda^2}\right) \approx -V'    .
\end{equation}
Given the above conditions, we also define our slow-roll parameters as follows, \\[5pt]
\begin{equation} \label{eps1}
\epsilon=\dfrac{1}{2}\dfrac{1}{(1+3C)} \left(\dfrac{V'}{V} \right)^2,
\end{equation} 
\begin{equation} \label{eps2}
\eta=\dfrac{V''}{V(1+3C)} ,
\end{equation} \\[5pt] where $C=K/\lambda^2$. We now impose the high-friction limit $C>>1$  \cite{YangFeiGaoGong2016}, which will enhance the damping introduced by the derivative coupling. The conclusion section presents a physical interpretation of the limit. In some sense, we are working a strong-coupling regime the value of the ratio $C$ will be assumed to be very large.\footnote{The presence of negative sign is not arbitrary. In the Eq. \eqref{eps1}, in the high friction limit the factor $(1+3C) \sim 3C$, if the sign of the coupling constant is positive that makes $\epsilon$ a negative quantity, therefore the negative sign is required.}
In this limit, equations Eq. \eqref{eps1} and Eq. \eqref{eps2} reduce to,
\begin{eqnarray}
\epsilon &=&\dfrac{\lambda^2}{6K} \left(\dfrac{V'}{V} \right)^2 ,\\
\eta &=&\dfrac{\lambda^2}{3K}\dfrac{V''}{V}.
\end{eqnarray}
The number of e-folds remaining before the end of inflation can be defined using the Eq. \eqref{second simplified eom} together with high-friction limit takes the form,
\begin{equation}
N=\int_{\phi_e}^{\phi} \dfrac{1}{\sqrt{2\epsilon}} \sqrt{\dfrac{3K}{\lambda^2}} d\phi \;.
\end{equation} 
We need the slow-roll parameters to be only a function of the potential, we need to express $K$ in terms of the potential. From Eq. \eqref{first simplified eom}, \eqref{second simplified eom}, \eqref{eps1} and \eqref{eps2} we have,

\begin{equation}
K=\dfrac{V}{6} \left( 1 \pm \sqrt{1-2\dfrac{V' \lambda^2}{V^3}}\right) \; .
\end{equation}
During slow roll, the first derivative of the potential varies slowly in comparison to the potential itself. From equation Eq. \eqref{first simplified eom} $V^2 \sim H^4$ toghether with the high friction limit, the quantity $\dfrac{ 2 V' \lambda^2}{V^3}$ is very small as compared to unity, so that  $K \approx \dfrac{V}{3}$. It is somewhat expected, given that $K=\dot{H}+H^2$, as during slow roll we already have the condition in the Hubble basis that $\dfrac{\dot{H}}{H^2} << 1$. We can now express all the slow-roll parameters in terms of the potential and the derivatives of the potential. The slow roll parameters take the simple form,  \begin{equation} \label{eps sim}
\epsilon=\dfrac{\lambda^2}{2V}\left(\dfrac{V'}{V} \right)^2 , \end{equation}

\begin{equation} \label{eta sim}
\eta=\dfrac{\lambda^2}{V} \dfrac{V''}{V} ,
\end{equation}
\begin{equation} \label{N}
N=\int_{\phi_e}^{\phi} \dfrac{1}{\sqrt{2 \epsilon}} \sqrt{\dfrac{V}{\lambda^2}} \; 	d\phi \;.
\end{equation} 
The scalar and tensor power spectra under the slow roll are respectively given by  \cite{Di_Marco_2024},
\begin{equation}
P_S(k) \sim \frac{1}{8\pi^2} \frac{H^2}{\epsilon_H} \bigg|_{k = aH} ,
\end{equation} and,
\begin{equation}
P_T(k) \sim \frac{2}{\pi^2} H^2 \bigg|_{k = aH}  ,
\end{equation} where $\epsilon_{H}=\dfrac{-\dot{H}}{H^2}$.
The spectral tilt ($n_s$) and the tensor to scalar ratio ($r$) take the following form
  \cite{Gao2025NonMinimalDerivativeCoupling} \cite{YangFeiGaoGong2016},
\begin{equation} \label{n_s}
n_s  \approx 1 - 8\epsilon + 2\eta,
\end{equation} 
\begin{equation} \label{r}
r \approx 16\epsilon \;.
\end{equation}

\section{Observational Constraints}
We will now validate the predicted results from theory with the latest ACT and Planck Data \cite{louis2025atacamacosmologytelescopedr6}.
The values of all the $\Lambda$CDM model parameters is listed in the appendix of the paper along with the values of $n_s$ and $r$. The data clearly shows the progressive increase in the values of $n_s$ as planck data is mixed and further increase when the lensing and BAO data are mixed \cite{louis2025atacamacosmologytelescopedr6}.

\subsection{Power Law Potential}
The simplest form of potential is the power-law potential \cite{Linde1983Chaotic} \cite{Linde1983ChaoticInflation}, which emerges as the leading-order terms in various kinds of potentials \cite{Martin2014Encyclopaedia}. It is useful in analytic calculation of the observables and validating with the observational values.
For the potential the,

\begin{equation} \label{power_law_potential}
V(\phi)=V_0 \phi^{n}, \end{equation} the parameters take the form as \cite{Gao2025NonMinimalDerivativeCoupling},
\begin{equation}
\epsilon=\dfrac{\lambda^2 n^2}{2 V_0 \phi^{n+2}} ,\label{eps_pl}
\end{equation}

 \begin{equation}
\eta=\dfrac{2(n-1)}{n} \epsilon  \;. \label{eta_pl}
\end{equation}
Inflation is going to end when any of the above slow-roll parameters becomes unity.
When $n$ ranges from 0 to 2 then $\eta < \epsilon$ and inflation will end according as $\epsilon(\phi_e)=1$. When exceeds 2, then $\eta > \epsilon$ and inflation will end according as $\eta(\phi_e)=1$. In the case when $0<n<2$, the value of $\phi_e$ such that $\epsilon$ becomes unity can be found using Eq. \eqref{eps_pl}. The the integration step in the Eq. \eqref{N} just involves integrating $\phi^{n+1}$ ,
 \begin{equation}
\phi_e = \left(\dfrac{\lambda^2 n^2}{2 V_0}\right)^{\dfrac{1}{(n+2)}} ,
\end{equation} 

\begin{equation}
N=\dfrac{V_0}{\lambda^2 n(n+2)} \tilde{\phi}^{n+2} - \dfrac{n}{2(n+2)} \;.
\end{equation} 
The second term is taken on the LHS, so that now RHS only depends on $\phi$ where $M=N+\dfrac{n} {2(n+2)}$ and $\tilde{\phi}$ is the value of the scalar field at horizon exit, leading to the equation in terms of $M$ , 
\begin{equation}
M=\dfrac{V_0}{\lambda^2 n(n+2)} \tilde{\phi}^{n+2} .
\end{equation}  
Similarly, when $n>2$,  \\
\begin{equation}
M = \dfrac{V_0 \tilde{\phi}^{n+2}}{n(n+2)\lambda^2} ,
\end{equation} where $M=N+\dfrac{(n-1)}{(n-2)} $ and $\tilde{\phi}$ is the value of the scalar field at horizon exit. Finally, using Eq. \eqref{n_s} and Eq. \eqref{r}, $n_s$ and $r$ take the following form, where $M$ takes the form depending if $0<n<2$ or $n>2$,

\begin{align}
n_s &= 1-\dfrac{2(n+1)}{M(n+2)}, \\
r &= \dfrac{8n}{(n+2)M} \;.
\end{align}
This is one of the few cases where we can perform a full analytical calculation. We see that for the exponent $n=1$, the theoretical prediction lies roughly on the outer boundary of 2$\sigma$ likelihood. However, for $n=1/3$, we notice that both the values lie in the acceptable range. All the other potentials are treated numerically. The step that actually requires numerical intervention is equation Eq. \eqref{N}, which involves performing the integration to find the suitable value of $\tilde{\phi}$. All other steps can, in principle, be done analytically. It is the recommended approach, especially for obtaining the expression of $\epsilon$ and $\eta$. The numbers that we are dealing with are very small; in these cases, numerical derivatives can lead to unexpected behaviour. The potentials are taken with parameters that we tune. For the numerical conversion of $C$ we have taken $\dfrac{\lambda^2}{V_0} = 10^{-4} $ \cite{Gao2025NonMinimalDerivativeCoupling}.
As a note, we will also show the effect of the NMDC term as compared to 
canonical inflation. For simplicity, take $n=1$ and $\dfrac{\lambda^2}{V_0}$ 
to be $10^{-4}$. Without the NMDC term, the equations,

\begin{equation}
    \epsilon_{sr} = \dfrac{1}{2} \left( \dfrac{V'}{V}\right)^2,
\end{equation}
where the subscript $sr$ denotes the parameter with respect to the minimal model and. Using Eq. \eqref{power_law_potential} for this particular case, setting 
$\epsilon_{sr}$ to be equals to 1 we obtain,

\begin{equation}
    \dfrac{1}{2} \dfrac{1}{\phi_e^2} = 1,
\end{equation}
so that $\phi_e=\sqrt{\dfrac{1}{2}} \simeq 0.707$. Here $\eta$ takes the form,

\begin{equation}
    \eta = \dfrac{V''}{V}.
\end{equation}
Using Eq. \eqref{power_law_potential}, $V''=n(n-1)\phi^{(n-2)}$; for $n=1$, 
$\eta$ becomes 0. The number of e-folds is then,
\begin{equation}
    N_{sr} = \int_{\phi_e}^{\tilde{\phi}} \dfrac{V}{V'} \, d\phi 
        = \int_{\phi_e}^{\tilde{\phi}} d\phi
        = \dfrac{\tilde{\phi}^2}{2}-\dfrac{\phi_e^2}{2} \;.
\end{equation}
Taking the same number of e-folds of 60 in both the cases, in slow roll case we have $\phi = \sqrt{120} \simeq 11 $ and $\epsilon_{sr} $ is 0.045. However in the NMDC case, setting $n=1$ in Eq. \eqref{eps sim} Eq. \eqref{eta sim} Eq. \eqref{N} we have $\phi \sim 6 \times 10^5$ and $\epsilon$ is 0.008 hence $n_s$ in the NMDC case is higher. 
\subsection{Exponential $\alpha$ Attractor}Exponential attractor \cite{Bhattacharya_2023} \cite{KalloshLinde2013} is a classic case of a potential that has the desired plateau as it approaches higher values. It has been is excellent agreement with earlier observational data. The slow-roll assumption is natual in this case,
\begin{equation}
V=V_0 \left[1 - a \exp \left( \, -\sqrt{\dfrac{2}{3 \alpha}}\phi\right)\right]
\end{equation} 
We then have the slow-roll parameters as follows:
\begin{align}
\epsilon
&=
\frac{\lambda^2 \exp\!\left( \sqrt{\frac{2}{3\alpha}}\, \phi \right)}
{3\alpha V_0 \left[ \exp\!\left( \sqrt{\frac{2}{3\alpha}}\, \phi \right) - 1 \right]^3}, \\[5pt]
\eta
&=
-\frac{2 \lambda^2 \exp\!\left( \sqrt{\frac{2}{3\alpha}}\, \phi \right)}
{3\alpha V_0 \left[ \exp\!\left( \sqrt{\frac{2}{3\alpha}}\, \phi \right) - 1 \right]^2}.
\end{align} 
We take $a=1$ for simplicity and $\alpha=0.008$, and then proceed to obtain the numbers. For $N=55$, we have the value of $n_s$ = 0.974 and $r$ = 0.025. For $N=50$ we get $n_s$ = 0.972 and $r$ = 0.028. For $N=60$ we have $n_s$ = 0.976 and $r$ = 0.023. All the values lie in the acceptable range. For values of $\alpha$ ranging from $10^{-5}$ to $10^2$, the values of $n_s$ are surprisingly stable; however, $r$ starts to increase, pushing it beyond the $2 \sigma$ agreement at $N=60$.
\subsection{Arc tan Potential}
The arctangent inflationary potential provides a smooth transition between a steep region and an asymptotically flat plateau, naturally supporting slow-roll inflation at large field values  \cite{Mukhanov2013},
\begin{equation}
V=V_0 \tan^{-1} a\phi  .
\end{equation}
Qualitatively, the behaviour of $\tan^{-1} \phi$ and the exponential $\alpha$ attractor is not so different. 
The slow roll takes the following form,
\begin{equation}
\epsilon = \dfrac{\lambda^2 a^2}{2 V_0 (1+a^2\phi^2)^2 (\tan^{-1} a\phi)^3} ,
\end{equation}
\begin{equation}
\eta=\dfrac{-2\lambda^2 a^3 \phi}{V_0 (1+a^2\phi^2)^2(\tan^{-1}a\phi)^2} \;.
\end{equation} \\[5pt]
For $N=60$, we get the value of $n_s$ = 0.978 and $r$ = 0.043. Similarly, for $N=50$ we have $n_s$ = 0.973 and $r$ = 0.05. In particular, the $r$ value is much higher compared to the former. TBased upon our numerical analysis we find that the most suitable value of $a$ = $ 0.4$, $a$ can take values from $10^{-3}$ to $10^2$ at $N=60$, which places it within a 2$\sigma$ range of the ACT observations.
\subsection{Hilltop Potential}
Hilltop inflation \cite{BoubekeurLyth2005} models are characterised by scalar field potentials in which inflation occurs near a local maximum, with the inflaton slowly rolling away from an unstable equilibrium. Such potentials naturally arise in scenarios with spontaneous symmetry breaking and effective field theories containing higher-order corrections,
\begin{equation}
V=V_0 \left(1-\dfrac{\phi^{p}}{\mu^p} \right).
\end{equation}
The slow roll parameter takes the following form,
\begin{equation}
\epsilon
=
\dfrac{
p^{2}\lambda^{2}\left(\frac{\phi}{\mu}\right)^{2p-2}
}{
2V_{0}\mu^{2}
\left[\,1-\left(\frac{\phi}{\mu}\right)^{p}\right]^{3}
},
\end{equation} 

\begin{equation}
\eta = -\dfrac{2(p-1)}{p} \left(\dfrac{1-\left(\frac{\phi}{\mu}\right)^p}{\left(\frac{\phi}{\mu}\right)^p}\right) \epsilon \;.
\end{equation}
We take $p=4$ \cite{Gao2025NonMinimalDerivativeCoupling} and $\mu=0.5$ for the calculation. For $N=60$ , $n_s$ takes the value 0.976 and $r$ = 0.02. At $N=50$, $n_s$ = 0.971 and $r$ = 0.03. $\mu$ can take the minimum value of 0.12 and arbitrary large value at $n=60$ to be in the 2$\sigma
$ agreement.
\subsection{Polynomial Attractor}Polynomial 
attractor \cite{Galante2015} \cite{Gao2025NonMinimalDerivativeCoupling} models form a well-motivated subclass of attractor inflation in which the inflaton potential is polynomial in the canonically normalised field but mapped through a curved field-space geometry  \cite{KalloshLinde2013}. Due to this geometric stretching, even simple polynomial potentials are transformed into plateau-like forms at large field values, leading to universal inflationary predictions largely insensitive to the detailed shape of the potential ,
\begin{equation}
V=V_0 \left(1-\dfrac{\mu^p}{\phi^{p}} \right)\;.
\end{equation}
 The slow-roll parameters are as follows,
\[
\epsilon
=
\frac{
p^{2} \lambda^{2} \left( \mu/\phi \right)^{2p+2}
}{
2 V_{0} \mu^{2}
\left[\,1 - \left( \mu/\phi \right)^{p} \right]^{3}
},
\]

\[
\eta
=
-\frac{
p(p+1) \lambda^{2} \left( \mu/\phi \right)^{p+2}
}{
V_{0} \mu^{2}
\left[\,1 - \left( \mu/\phi \right)^{p} \right]^{2}
}
=
-\frac{2(p+1)}{p}\,
\frac{
1 - \left( \mu/\phi \right)^{p}
}{
\left( \mu/\phi \right)^{p}
}\,
\epsilon \;.
\]
Here we have taken $\mu=0.1$ and $p=3$ and at $N=60$ we have $n_s$ = 0.976 and $r$ = 0.01. At $N=50$, we have a value of $n_s$ = 0.971 and and r = 0.014. The range of values that $\mu$ can take in this case is $10^{-5} < \mu$ at $N=60$ to be in $2\sigma$ agreement of the ACT observations. 
\begin{figure}[H]
\centering
\includegraphics[scale=0.7]{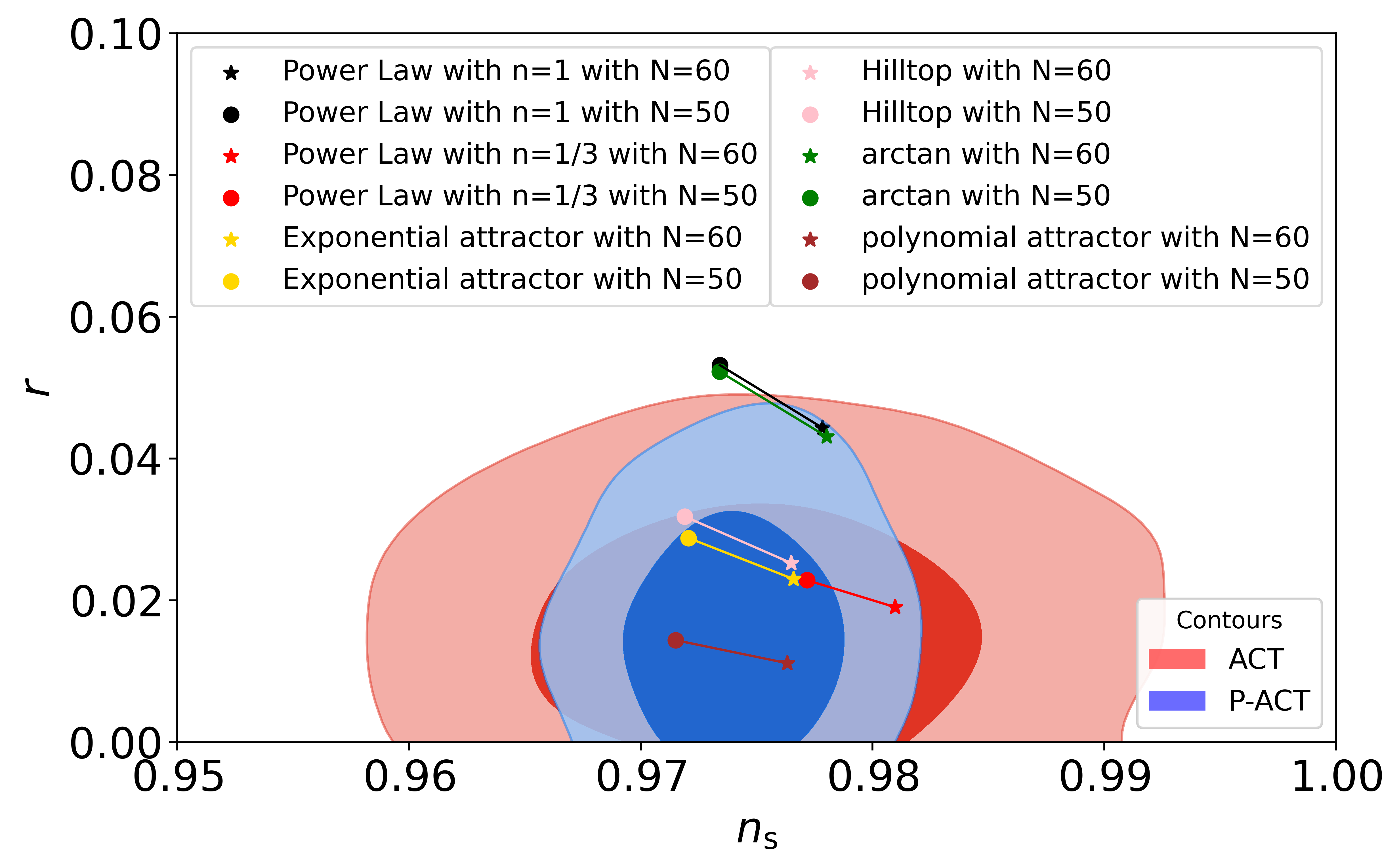}  
\caption{1-$\sigma$ (dark) and 2-$\sigma$ (light) likelihood contours in the 
$n_s-r$ plane. The \textcolor{red}{red} contours correspond to ACT\,+\,BK (BICEP/Keck)  \cite{BICEPKeck:2021gln} 
constraints and the \textcolor{blue}{blue} contours correspond to 
Planck\,+\,BK constraints  \cite{villagra2025atacamacosmologytelescopehighredshift}
 \cite{louis2025atacamacosmologytelescopedr6} \cite{calabrese2025atacamacosmologytelescopedr6}. 
The theoretical predictions of the NMDC model are shown for the power-law 
($n=1$ and $n=1/3$), exponential $\alpha$-attractor, Hilltop, 
arctan, and polynomial attractor potentials. Dots represent $N=50$ and 
asterisks represent $N=60$ e-folds.}
\label{fig:ns_r}
\end{figure}
We can clearly see that the model agrees with the observation within the $2\sigma$ estimate. The power law potential and the arctan potential show marginal deviation. For all the potentials, the higher $N$ value is always lower in $r$ and higher in $n_s$ as expected. Thus, within the framework discussed, the extra damping brings them closer to the observation. The power law potential with $n=1$ lies entirely outside both the Planck and ACT contours, with $r \sim 0.053$ at $N=50$, placing it in strong tension with current data, and while increasing $N$ reduces $r$ somewhat, neither value achieves consistency at even the $2\sigma$ level. The power law potential with $n=1/3$ fares better, with both $N=50$ and $N=60$ sitting near the right edge of the outer ACT $2\sigma$ contour at $n_s \sim 0.980$, making it marginally consistent though requiring higher e-fold counts to avoid tension. The exponential and Hilltop potentials both place their predictions comfortably within the ACT $1\sigma$ region at $n_s \sim 0.972$-$0.975$ and $r \sim 0.025$-$0.032$, benefiting from their naturally flat plateaus at large field values which suppress the tensor-to-scalar ratio. The arctan potential, despite its plateau-like shape, produces $r \sim 0.044$--$0.053$ that tracks closely with the power law $n=1$ predictions and remains in mild tension with the data even at $N=60$. Finally, the polynomial attractor potential provides the best agreement with ACT DR6, with both e-fold values falling deep within the inner $1\sigma$ contour at $r \sim 0.012$--$0.015$, owing to its attractor-class behaviour.

\section{Conclusion and Discussion}
We started by introducing a non-minimal derivative coupling (NMDC) term in the high-friction limit in order to reconcile the increase in the spectral tilt suggested by the recent ACT+BAO data. The presence of the NMDC term effectively enhances the friction experienced by the inflaton, thereby modifying its background evolution and allowing for deviations from the predictions of canonical single-field inflation. Working within the slow-roll framework, and motivated by the canonical theory, we defined the slow-roll conditions and parametrised them in terms of the usual slow-roll parameters 
$\epsilon$ and $\eta$. In this regime, the high-friction limit simplifies the dynamics while still capturing the essential effects of the derivative coupling on the inflationary observables. Equations Eqs. \eqref{eps sim}, \eqref{eta sim} and  \eqref{N} clearly indicate that the high-friction factor reduces the values of their respective counterparts thereby increasing the sample size of possible potentials.
Using Eqs. \eqref{eps sim}, \eqref{eta sim} and \eqref{N}, we can relate the slow-roll parameters to their cannonical counterparts denoted by the subscript $sr$,
\begin{equation} \label{eps_rel}
    \epsilon = \dfrac{\lambda^2}{V} \epsilon_{sr}
\end{equation}

\begin{equation} \label{eta_rel}
    \eta =  \dfrac{\lambda^2}{V} \eta_{sr}
\end{equation}

\begin{equation} \label{N_rel}
    \dfrac{dN}{d\phi} = \dfrac{V}{\lambda^2} \dfrac{dN_{sr}}{d\phi}
\end{equation}
From Eqs. \eqref{eps_rel} and \eqref{eta_rel}, we can see that for any given potential the modified values will admit smaller values of the mode exit as compared to their respective counterparts. Moreover, along with Eq. \eqref{N_rel}, it also translates into smaller values at the mode exit value of $\phi$, reducing the total sum, which is subtracted from unity for the calculation of the spectral tilt.

In summary, NMDC scenario introduces new dynamical freedom due to the non-minimal interaction term coupling the scalar field to gravity. It influences the effective friction coefficient acting on the inflaton field, thus affecting the inflationary evolution. Furthermore, NMDC can render potential solutions to inflation viable that would otherwise be invalid under the observational constraints of minimalist approaches like the chaotic potential\cite{Linde1983ChaoticInflation}. The expanded parameter space enables a broader range of viable theoretical models to be explored, which, in its absence, remain unobservable.

Figure~\ref{fig:ns_r} shows the ACT-lite$+$BK and Planck$+$ACT-lite$+$BK datasets, overlaid with the theoretical 
predictions of seven inflationary potentials evaluated at $N=50$ and $N=60$ e-folds. 
The polynomial attractor model is the most favoured, with both $N=50$ and $N=60$ predictions lying well within the $1\sigma$ region of both datasets. The exponential and hilltop potentials are also consistent at the $1\sigma$ level, placing them among the viable candidates. The power-law potential with $n=1/3$ is marginally consistent at $2\sigma$ for $N=60$, though its $N=50$ prediction is disfavoured. The arctan potential lies at or beyond the $2\sigma$ boundary for both datasets and both e-fold values, rendering it observationally disfavoured. The power-law potential with $n=1$ is strongly disfavoured, lying well outside the $2\sigma$ contours of both datasets. Overall, the results favour inflationary models predicting a low tensor-to-scalar ratio $r \lesssim 0.03$ at the observed spectral index $n_s \approx 0.97$--$0.98$, strongly disfavouring large-field power-law inflation.

The model is applicable with the central assumption of slow-roll inflation. Without slow-roll, the analysis done in Eq.\eqref{higer_order} is not valid. Compared to the coupling with the Einstein tensor \cite{Gao2025NonMinimalDerivativeCoupling}, the higher derivatives do not exactly vanish. Interestingly, the form of observational variables remains the same.

The theoretical predictions obtained from this analysis suggest that NMDC-based models offer a promising approach to explaining the observed spectral properties of primordial perturbations. In particular, the modified slow-roll dynamics lead to an enhancement of the scalar spectral tilt, bringing it closer to the values preferred by current observations. This suggests that derivative couplings to curvature can play a non-trivial role in shaping inflationary predictions without requiring drastic modifications to the underlying inflationary framework. There remains considerable scope for extending this work. Additional curvature-dependent terms or alternative forms of derivative couplings can be incorporated, which were previously held back by the issue of ghosts, introducing new parameters that may further improve agreement with observational data. As we have shown that in purview of the slow-roll regime, the higher derivative terms can be dropped. For instance, derivative couplings to the full Riemann tensor $R^{\mu\nu\alpha\beta}\partial_\mu\phi\,\partial_\alpha\phi$ generically introduce higher-order equations and Boulware--Deser ghosts \cite{Amendola1993} \cite{BoulwareDeser1972}, unlike the special Einstein tensor contraction studied here. Similarly, the Weyl-squared correction $\alpha\, C_{\mu\nu\rho\sigma}C^{\mu\nu\rho\sigma}$ produces ghost modes in the scalar sector that destabilize the inflationary background \cite{DeFelice2023Weyl}. Kinetic-dependent Gauss--Bonnet couplings $f(\phi,X)\,\mathcal{G}$, which belong to the GLPV class, also face gradient instabilities unless specific degeneracy conditions are satisfied \cite{Shahidi2019}. Exploring such extensions would allow a more systematic understanding of how different NMDC structures influence both background dynamics and perturbations. It will also be interesting to investigate whether a more general framework, encompassing a wider class of curvature couplings, can provide testable predictions in light of upcoming high-precision cosmological observations.

\end{document}